\documentclass[useAMS,usenatbib]{mn2e}
\usepackage{graphicx,amssym}
\usepackage{color}
\newcommand{\comment}[1]{}

\newcommand{\bc}{\begin{center}}
\newcommand{\ec}{\end{center}}

\title[A modified star formation law]
      { A modified star formation law as a solution to 
open problems in galaxy evolution
}

\author[L.Wang, S.M.Weinmann, E.Neistein]
       {Lan Wang$^{1}$$\thanks{Email: wanglan@bao.ac.cn}$,
        Simone M. Weinmann$^{2}$,
        Eyal Neistein$^{3}$
	\\
        $^1$Partner Group of the Max Planck Institute for Astrophysics, National Astronomical Observatories, \\
        Chinese Academy of Sciences, 
  20A Datun Road, Chaoyang District, Beijing, China\\
	$^2$Leiden Observatory, Leiden University, P.O. Box 9513, 2300 RA Leiden, 
  The Netherlands  \\
	$^3$Max-Planck-Institute for Extraterrestrial Physics, Giessenbachstrasse 1, 
   85748 Garching, Germany        }
\begin{document}

\date{Accepted 2011 ???? ??. 
      Received 2011 ???? ??; 
      in original form 2011 ???? ??}

\pagerange{\pageref{firstpage}--\pageref{lastpage}} 
\pubyear{2011}

\maketitle

\label{firstpage}

\begin{abstract}

In order to reproduce the low mass end of the stellar mass function,
most current models of galaxy evolution invoke very efficient supernova 
feedback. This solution seems to suffer from several shortcomings however, like
predicting too little star formation in low mass galaxies at z=0.
In this work, we explore modifications to the star formation (SF) law
as an alternative solution to achieve a match to the stellar mass function.
This is done by applying semi-analytic models based on De Lucia \& Blaizot, but
with varying SF laws, to the Millennium and Millennium-II simulations, within
the formalism developed by Neistein \& Weinmann.
Our best model includes lower SF efficiencies than predicted by the 
Kennicutt-Schmidt law at low stellar masses, no sharp threshold of cold gas 
mass for SF, and a SF law that is independent of cosmic time. These simple 
modifications result in a model that is more successful than current standard 
models in reproducing various properties of galaxies less massive than 
$10^{10}{\rm M}_{\odot}$. The improvements include a good match to the 
observed auto-correlation function of galaxies, an evolution of the stellar
mass function from $z=3$ to $z=0$ similar to observations, and a better 
agreement with observed specific star formation rates. However,
our modifications also lead to a dramatic overprediction of the cold mass 
content of galaxies. This shows that finding a successful model 
may require fine-tuning of both star formation and supernovae feedback, 
as well as improvements on gas cooling, or perhaps the inclusion of a yet 
unknown process which efficiently heats or expels gas at high redshifts.

\end{abstract}


\section{Introduction}
\label{sec:intro}

Ever since first introduced by  \citet{white1991}, semi-analytic models 
of galaxy formation and evolution  \citep[hereafter SAM;][]{kauffmann1999, 
kang2005, croton2006, Bower2006, cattaneo2006, DLB2007, monaco2007,
somerville2008, Khochfar2009} have been successfully used to study how 
different physical processes determine the formation and evolution of galaxies.
Based on halo merger trees extracted from $N$-body simulations or 
analytic methods, SAMs follow the main processes that are thought to 
affect the properties of galaxies, like gas cooling, star formation, feedback, 
and merging. These models provide a useful tool to study the interplay and 
the relative importance of these different physical processes. A detailed 
review of the semi-analytic method can be found in \citet{baugh2006}.

Although various observational properties of galaxies are matched by the 
models, current SAMs still have problems in reproducing some important 
observations and up to now, there is no semi-analytic model that 
is able to fit all the key statistical properties of the observed
galaxy population. 
For example, \citet{guo2010}  show that the model of 
\citet{DLB2007} substantially overproduces the low mass end of the stellar
mass function of galaxies. This problem becomes more severe as the resolution
of the underlying dark matter simulation increases. In addition,
stellar mass functions at high redshifts are normally not well reproduced 
\citep{fontanot2009, marchesini2008}; 
a good match to the 2-point auto-correlation function of
galaxies is so far difficult to achieve  \citep[e.g.][]{guo2010}; 
and the relation between the specific star formation rate and galaxy
stellar mass deviates from observations  \citep{somerville2008, fontanot2009}.

These discrepancies may have various reasons: inaccurate physical
modeling of the processes that govern galaxy formation; technical 
problems in tuning the model against a large set of observational constraints; 
the use of a fixed functional form for a poorly understood 
process, which overly limits the freedom in tuning the model;
or a wrong cosmological model adopted in the simulations. This large range
of possibilities makes it difficult to correct identified discrepancies between
model and observations.

For example,  \citet{guo2010} tried to fix the over-prediction of the low mass 
end of the stellar mass function, as modeled by  \citet{DLB2007}, by increasing 
the effect of supernovae feedback. They managed to reproduce the amplitude 
of galaxy stellar mass function, and also obtained a reasonable match to 
the galaxy luminosity functions in different bands. However, they also 
predict a too large fraction of red galaxies at low masses, too high
amplitudes of the stellar mass functions in the redshift range 
of [0.8, 2.5], and galaxy auto-correlation functions that are too high 
for galaxies less massive than $6\times10^{10}{\rm M}_{\odot}$.
Moreover, the high feedback efficiency as used by  \citet{guo2010} is 
physically difficult to motivate  \citep{benson2003}, and is 
far more efficient than various solutions adopted by hydrodynamical 
simulations  \citep{maclow1999, strickland2000, avilareese2011}.

In this work, we therefore explore an alternative solution.
We tune the SF recipe instead of the supernovae (SN) feedback
to study how our changes affect different statistics of galaxies, 
and to what degree the discrepancies mentioned above can be alleviated.
Most of the current SAMs use an analogue to the empirical 
Kennicutt--Schmidt 
law \citep{schmidt1959, kennicutt1998} to calculate the star formation 
rate (SFR) in galaxies. In this standard prescription, the SFR is roughly 
proportional to the cold gas mass, and scales inversely with the typical 
time-scale of a galactic disk \citep[e.g.][]{cole2000}. This law is 
combined  with a sharp threshold at low gas densities, below which no SF 
occurs  \citep{kauffmann1996, croton2006}. The use of a such a threshold
is motivated both theoretically \citep[e.g.][]{toomre1964,kennicutt1989, 
schaye2004} and observationally \citep[e.g.][]{martin2001}. 

This simple SF law is however likely an oversimplification.
In recent years observational determinations of the SF law in galaxies have 
become increasingly refined, using various gas components (HI, CO) in 
combination with more reliable estimates of SF,
based on UV and IR light. 
These recent findings can be summarized as follows: First, there are 
indications that the threshold for SF at low mass densities
is not sharp. Instead, the SF efficiency drops off as a steep power-law at 
low gas densities \citep{kennicutt2007, bigiel2008, wyder2009, 
roychowdhury2009,bigiel2010}. Second, several studies find that
the SF rate is correlated more strongly with the mass of molecular gas (H$_2$) 
than with the the atomic gas \citep{wong2002,bigiel2008,leroy2008},
probably also at high redshift \citep{bouche2007,genzel2010}. 
This shows that simply correlating the star formation rate with the total cold
gas density in models may not always lead to realistic results. It is 
also not clear
which gas mass correlates best with the SF rate when averaging over 
the entire galaxy \citep[e.g.][]{saintonge2011}. 
Third, there are both theoretical and observational indications 
that the normalization of the SF law might be lower at high redshift than 
locally \citep[e.g.][]{wolfe2006, rafelski2010, gnedin2010, agertz2011, 
krumholz2011}. This means that simply extrapolating the local relation, 
as done in most SAMs, may be incorrect.

Several recent  models have attempted to address these issues. 
\citet{baugh2005}, \citet{weinmann2011a}, and \citet{krumholz2011} 
specifically lower the quiescent star formation efficiencies at high 
redshifts in order to better match some observed properties of high 
redshift galaxies. Both  \citet{fu2010} and  \citet{lagos2010, lagos2011} focus
on the first two of the above points, and present SAMs with updated and 
much more detailed SF recipes in comparison to previous SAMs. They do not 
include a sharp threshold for SF, and their SF rate depends on the molecular 
gas density instead of on the cold gas mass, following recent 
empirical and theoretical models \citep[e.g.][]{blitz2006, krumholz2009a}.
Their detailed models for SF depend on various internal properties 
of the disk, like size and pressure, which are not trivial to 
model in a SAM.

It is certainly worthwhile to try and
include more detailed and observationally 
and theoretically better motivated SF law into SAMs. We present a 
complementary approach in this work, without taking into account
complicated processes on sub-galactic
scales, like the conversion from atomic to molecular gas. 
We instead try to solve in a straightforward way the inverse problem, 
namely which realistic SF law at galactic scales is required to improve the
agreement between model galaxies and observations.
We start with a standard model, similar to \citet{DLB2007} and 
Neistein \& Weinmann (2010), and assume that the SF rate 
depends on the cold gas mass in the galaxy, cosmic time, and in addition 
the host halo mass. We make simple change to the standard SF law that are 
qualitatively, but not quantitatively plausible, and explore how those 
impact on the properties of galaxies. We show that we
can improve the agreement between 
SAMs and observations in several key aspects in this way.

We use the model developed by Neistein \& Weinmann (2010), and implement
it on both the Millennium Simulation  \citep[MS,][]{springel2005} 
and the Millennium-II Simulation  \citep[MS-II,][]{boylan2009}. 
The properties of galaxies with masses as low as $\sim$ $10^{8}{\rm M}_{\odot}$
can probably be studied reliably with the help of these simulations  
\citep{guo2010}, although this can be resolution dependent for extreme 
models (see Neistein \& Weinmann 2010). Our aim is to 
investigate how much a model based on \citet{DLB2007} can be changed and
improved by tuning the SF law alone. We thus do not
change gas cooling and feedback in our models, but leave it at the
default standard values (except in one case, 
for illustration, as will be explained).

This paper is organized as follows. In section \ref{sec:model} we present
the different models used in this work. Our starting point is a model
that is based on the widely used SAM of \citet{DLB2007}, with an improved 
prescription for hot gas stripping of satellite galaxies. We then develop 
four different models. Of these, models 2 and 3 include the key changes to 
the quiescent and burst mode star formation. In 
section \ref{sec:results} we present predictions for the stellar mass
functions at low and high redshifts, the relation between galaxy specific
SF rate and stellar mass, the galaxy cold gas mass function, the 
auto-correlation functions, and the SF rate density as a function of 
redshift. A discussion of the implications of our results, 
and the conclusions are presented in section \ref{sec:conclusion}.


\section{models}
\label{sec:model}

The semi-analytic models presented in this paper are applied to both the
MS and MS-II simulations. The cosmological parameters in the simulations 
are consistent with a combined analysis of the 2dFGRS  \citep{colless01} 
and the first year WMAP data  \citep{spergel03}, with $\Omega_{\rm m}=0.25$, 
$\Omega_{\rm b}=0.045$, $h=0.73$, $\Omega_\Lambda=0.75$, $n=1$, and $\sigma_8=0.9$.
Note that these parameters are different from  the latest WMAP 7 
year results. Both simulations follow $N= 2160^3$ particles from redshift 
$z=127$ to the present day. The MS has a particle mass resolution of 
$8.6\times10^{8}\,h^{-1}{\rm M}_{\odot}$, with a comoving box of $500\, h^{-1}$Mpc 
on a side. The MS-II has a mass resolution of $6.9\times10^{6}\,h^{-1}{\rm M}_{\odot}$,
 with a box of side $100\, h^{-1}$Mpc.

\citet{NW2010} (hereafter NW2010) developed a new formalism for modeling
galaxy formation and evolution, which is similar to the standard SAMs,
except that the efficiencies of processes like gas cooling, star
formation and feedback are assumed to depend only on the host halo
mass and cosmic time. NW2010 have shown that this new method
produces a very similar population of galaxies like 
standard SAMs. The method is simple and flexible, which makes it easy 
to change recipes in order to fit selected observational constraints.

All the models within this work are based on a simple set of differential 
equations, that follow the mass of gas and stars within galaxies. We adopt 
the model of \citet[][hereafter DLB07]{DLB2007} as our starting point, as 
was done in NW2010. The reader is referred to these papers, and 
to \citet{croton2006}, for more details on the model assumptions.
Here we highlight a few features that will be important for the discussion 
below. The SF law is assumed to be:
\begin{equation}
{\dot{M}_{\rm star}} = {f_s}(M_{\rm cold}-M_{\rm crit}) 
\label{eqn:2}
\end{equation}
$f_s$ is the SF efficiency in units of Gyr$^{-1}$, and $M_{\rm crit}$ is the 
critical mass of cold gas, below which no star formation 
occurs \citep{kennicutt1998}.
Satellite galaxies are followed along with their host subhaloes. 
Once the subhaloes are stripped and cannot be identified anymore, we compute 
the radial distance, $r_{\rm sat}$, between the satellite and the central 
subhalo within the group. We then allow the satellite galaxy to spiral 
in further, and estimate the time it merges into the central object 
by using dynamical friction estimate:
\begin{equation}
t_{\rm df} = \alpha_{\rm df} t_{\rm C} = \alpha_{\rm df} \frac{1.17 V_{\rm v} r_{\rm sat}^2}{G M_{1} \ln \left( 1 + M_{\rm h}/M_{1}\right)} \, .
\label{eqn:tdf}
\end{equation}

Here $t_{\rm C}$ is the Chandrasekhar estimate for the dynamical friction 
timescale, where $V_{\rm v}$ is the virial velocity of the central subhalo, 
$M_{\rm h}$ is its mass, and $M_1$ is the baryonic (cold gas and stellar) mass 
of the satellite galaxy. $\alpha_{\rm df}$ describes the ratio of the adopted 
dynamical friction time over the Chandrasekhar estimate. 
When galaxies finally merge we assume a SF burst of the type:

\begin{equation}
{\dot{M}_{\rm star,\rm burst}} = {{\alpha}_{\rm burst}}{(M_{1,\rm cold}+M_{2,\rm cold})} \,,
\label{eqn:5}
\end{equation}
with
\begin{equation}
{\alpha_{\rm burst}} = {0.56}(M_1/M_2)^{0.7} \,.
\label{eqn:6}
\end{equation}
Here $M_1$ and $M_2$ denote the baryonic mass in the merging galaxies.
Following \citet{croton2006}, this formula is derived from fitting the 
results of hydrodynamical simulations \citep[e.g.][see section 2.4 for more
details]{cox2004}.

In the following subsections, we describe all the models used in this paper in
detail.

\subsection {Model 0}

Our model 0) is very similar to model 0) in NW2010, and should thus be also 
similar to DLB07.
For the SF law above (Eq.~\ref{eqn:2}) we use the same fitting
functions as in NW2010:
\begin{equation}
f_s = 2.04 \,M_{12}^{0.094}10^{-0.039[\log M_{12}]^2}t^{-0.82} \,,
\label{eqn:3}
\end{equation}
and
\begin{equation}
M_{\rm crit} = 0.36 \, f_s^{-1} M_{12}^{0.68} t^{-0.52} \,.
\label{eqn:4}
\end{equation}
Here $t$ is the Hubble time in units of Gyr, and $M_{12}=M_{\rm halo}/10^{12}$ 
is the halo mass in unit of $10^{12}h^{-1}{\rm M}_{\odot}$.

There are a few minor modifications made here in comparison to NW2010, 
that are related to the extended range in halo mass we use here 
(a minimum halo mass of $\sim10^8$ h$^{-1}\,M_\odot$ in comparison
to $\sim10^{10}$ h$^{-1}\,M_\odot$  in NW2010). For more details on how 
we extend the recipes from NW2010 to low mass haloes, the reader is 
referred to Appendix A. Fig.~\ref{fig:fiducial} shows that the 
amplitude of the low mass end of the stellar mass function (SMF) for 
model 0) is comparable to the DLB07 result when 
applied to MS-II simulation.

\subsection {Model 1}
Model 1) is the fiducial model used in this work. It is based on model 0)
as presented in the previous subsection, but includes two further changes: 
the hot gas stripping of satellite galaxies is slowed down considerably 
compared to DLB07 following  \citet{weinmann2010}, and a larger dynamical 
friction time is assumed.

In the DLB07 model, the hot gas component of a galaxy is stripped completely 
once it falls into a larger group and becomes a satellite. Satellite galaxies 
subsequently consume their remaining cold gas due to star formation and 
efficient SN feedback, and the SF rate ceases on a short timescale of 1 - 2 Gyr.
This leads to most satellite galaxies displaying red colour, which is in 
contradiction with observations \citep{wang2007, weinmann2009}. Following
the model suggested by  \citet{weinmann2010}, we assume in model 1)
that the hot gas component of satellite galaxies decreases at the same rate 
as their surrounding dark matter haloes, which lose mass due to tidal stripping.
This treatment provides a physically better motivated description of the 
behaviour of the hot gas component of satellite galaxies and improves 
agreement with observations  \citep{weinmann2010}. A similar model is 
included in the recent SAM of \citet{guo2010}.

The second change with respect to model 0) is that the parameter 
$\alpha_{\rm df}$, which describes the ratio of dynamical friction time for 
galaxy mergers over the Chandrasekhar formula (Eq.~\ref{eqn:tdf} above), 
is set to $5$, in contrast to the value $2$ as adopted in model 0) 
and in DLB07. This corresponds to a larger time scale for satellite galaxies 
to merge with the central galaxy, and is chosen in order to get 
auto-correlation functions in better agreement with observations for all 
the models (see below). Up to now there is no solid consensus on what
$\alpha_{\rm df}$ should be in models \citep{boylan2008, jiang2008, mo2010}.

\begin{figure}
\bc
\hspace{-0.4cm}
\resizebox{8cm}{!}{\includegraphics{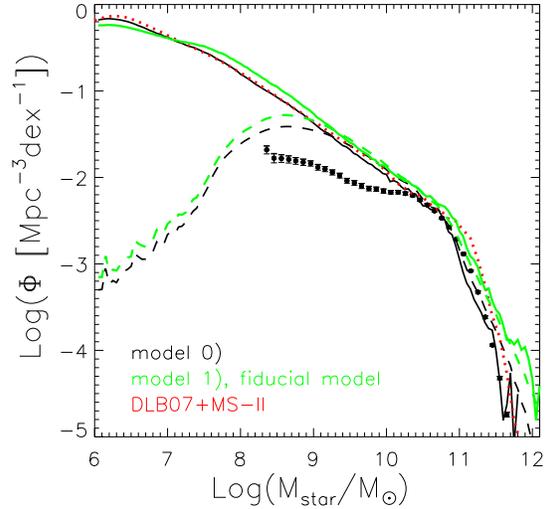}}\\%
\caption{The stellar mass functions of model 0) (black lines) and the fiducial
model 1) (green lines). Dashed lines are obtained using the MS simulation, and
solid lines are for the MS-II simulation. The red dotted line shows the 
stellar mass function using the DLB07 SAM for MS-II simulation. Filled dots 
with error bars show the observed stellar mass function of
SDSS DR7 result \citep{li2009,guo2010a}.
}
\label{fig:fiducial}
\ec
\end{figure}

Green lines in Fig.~\ref{fig:fiducial} show the SMF of the fiducial model 1), 
with dashed and solid lines for results when applied to MS and MS-II 
simulations. Model 1) SMF are in general slightly higher than model 0). 
This is mainly due to the slower stripping of hot gas in satellites adopted 
in model 1). Retaining their hot gas reservoir for longer, satellite
galaxies can continue forming stars for a considerable time and thus end 
up with a higher stellar mass than in model 0). When merging into central 
galaxies, they also add more mass to their centrals. Although our choice of 
a larger dynamical friction time, with $\alpha_{\rm df}=5$, delays merger to 
some degree and thereby decreases the amount of mass added to centrals, 
this effect is apparently smaller than that of the modification to the hot gas 
stripping. When comparing results of model 0) and model 1) for the MS 
and MS-II simulations, the SMF in the MS exceed the SMF in the MS-II
at intermediate masses. This is not the case for DLB07  \citep{guo2010}.
This excess may be related to the fact that our extension of the cooling
efficiencies do not exactly match those used for DLB07 in  \citet{guo2010}.

\subsection {Model 2}
In model 2), we make modifications to the SF law in the quiescent mode,
and keep all the other components of the model exactly the same as in the
fiducial model 1). The modifications include: a) $M_{\rm crit}=0$, 
which means no threshold cold gas mass for SF;
b) the SF efficiency depends more strongly on halo mass;
c) the SF efficiency does not depend on the Hubble time.

The SF law in model 2) is chosen such  that the low mass end slope of SMF 
in the model is comparable to the observed SDSS result, when applied to MS 
simulation. This results in
\begin{equation}
M_{\rm crit}=0 \,,
\label{eqn:7}
\end{equation}
\begin{equation}
{f_s} = 0.41 \, {M_{12}^{0.94}10^{-0.30[\log M_{12}]^2}} \,.
\label{eqn:8}
\end{equation}

\begin{figure*}
\bc
\hspace{-1.4cm}
\resizebox{16cm}{!}{\includegraphics{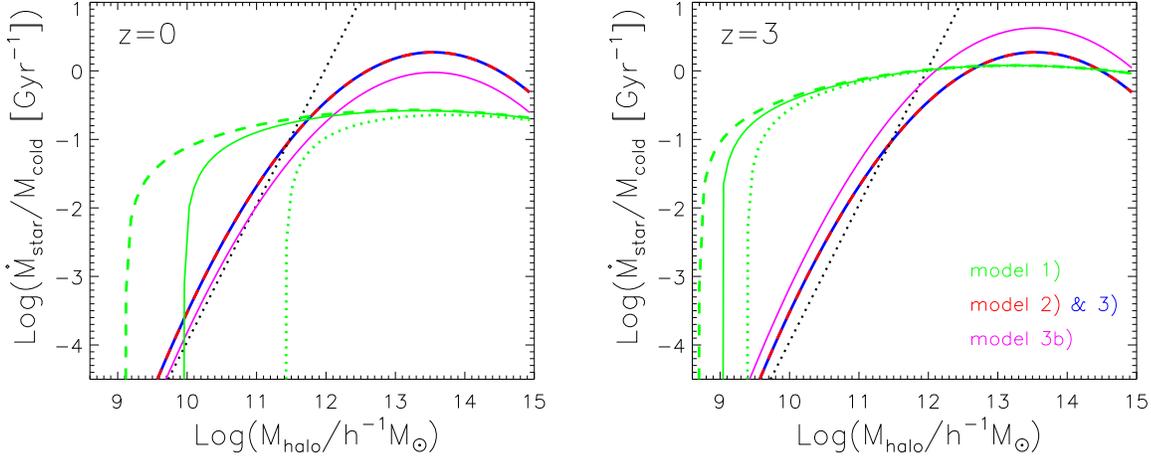}}\\%
\caption{
Star formation efficiencies in the quiescent mode as a function of halo 
mass at $z=0$ (left panel) and $z=3$ (right panel). Green lines are 
the efficiencies in model 1). Dashed, solid and dotted green lines are for 
the cases when the cold gas--halo mass ratios are [84, 50, 16] percentile 
values in model 1), which are [$0.13$, $0.04$, $0.007$] at $z=0$, and are 
[$0.14$, $0.09$, $0.05$] at $z=3$. Red-blue line is the efficiency of star 
formation for model 2) and 3). The magenta lines 
are for efficiencies in model 3b). Black dotted line shows a power law 
with index 2 for reference.
}
\label{fig:SFR}
\ec
\end{figure*}

Fig.~\ref{fig:SFR} compares the SF efficiency (SFE, defined as the ratio
between SF and the mass of the cold gas) in the quiescent mode in
model 1) and model 2), as a function of halo mass at two redshifts. Green
lines are efficiencies in model 1). Solid green lines are for a 
cold gas--halo mass ratio of $0.04$/$0.09$ at $z=0$/$z=3$, which
corresponds to the median cold gas--halo mass ratio in model 1).
Dashed green lines are for a higher cold gas--halo mass ratio of
$0.13$/$0.14$ at $z=0$/$z=3$, and dotted green lines are for
a lower ratio of $0.007$/$0.05$ at $z=0$/$z=3$.
These values correspond to the 16th and 84th percentiles of the 
distributions in cold gas masses in model 1). The SF efficiency in model 2) 
is shown as red-blue lines, which follow a power law of index $2$ 
(plotted as black dotted lines in Fig.~\ref{fig:SFR}) for haloes less 
massive than $\sim 10^{11.5}h^{-1}{\rm M}_{\odot}$. Compared to model 1), model 2) 
has a lower SF efficiency at $z=0$ for halo masses between $10^{10}$ and 
$\sim 10^{11.5}h^{-1}{\rm M}_{\odot}$. At $z=3$, the SF efficiency is much lower in 
model 2) than in model 1) for haloes less massive than 
$\sim 10^{12}h^{-1}{\rm M}_{\odot}$.

\subsection {Model 3}

When applying model 2) to the MS simulation, the amplitude of the SMF
at the low mass end decreases dramatically and matches the observation
of SDSS (red dashed line in Fig.~\ref{fig:SMF}). However, when applied
to the MS-II simulation, the low mass end of SMF is still too high (red 
solid line in Fig.~\ref{fig:SMF}). We have tested that even when truncating
all SF in quiescent disk mode in haloes less massive than 
$\sim 10^{11.5}h^{-1}{\rm M}_{\odot}$,
the problem still exists. This is because galaxy mass grows not only by 
quiescent SF in disks, but also by merger-induced bursts. These become 
more significant if the quiescent SF efficiency is decreased, due to the 
resulting increase in the cold gas masses.

In model 3), we therefore further modify the SF law in the burst mode, while 
keeping the same SF law for the quiescent mode as in model 2).
The star burst efficiency used in current SAMs 
(Eq.~\ref{eqn:6}) is derived from fitting the results of 
hydrodynamical simulations for mergers of mass ratios ranging from 1:10 to 
1:1 \citep{cox2004, mihos1994, mihos1996}. The
parameters in those simulations are set to make the SF
in an isolated disk galaxy consistent with the Kennicutt-Schmidt law.
As the quiescent SF of dwarf galaxies themselves is lower
than that predicted by the Kennicutt-Schmidt law, it is quite possible
that those simulations over-predict the burst efficiency for low mass
galaxies. Therefore we introduce a halo mass dependence for the
efficiency of merger-induced bursts in model 3), and make
it inefficient for low mass haloes.

For haloes less massive than $M_0$, we modify the burst efficiency to be
\begin{equation}
{\alpha_{\rm burst}} = {0.56}(M_1/M_2)^{0.7}{\times}({M_{\rm halo}}/{M_0})\,.
\label{eqn:9}
\end{equation}
while it remains unchanged for higher mass haloes. We use
$M_0 = 10^{11.5}h^{-1}{\rm M}_{\odot}$. This critical halo
mass is empirically determined, and roughly corresponds to 
the mass where the relation between galaxy stellar mass and halo mass changes 
its slope \citep[e.g.][]{wang2006,moster2009}, and galaxy 
formation efficiency reaches its maximum \citep{guo2010a}. 
Physically, SN feedback and reionization are
believed to cause a low galaxy formation efficiency at
low halo masses, and AGN feedback may be responsible for the low efficiency 
in high mass haloes \citep{guo2010a}. The combined effect of these mechanisms
may be weakest at this critical halo mass, explaining the peak in galaxy 
formation efficiency in the current theory. Therefore
we choose this value as the threshold below which the SF efficiency is 
assumed to be suppressed. The typical stellar mass
of galaxies that reside in those haloes is about $10^{10.5}{\rm M}_{\odot}$, 
below which model 0) and model 1) predict too many galaxies and thus too much
SF. By decreasing the SF 
for galaxies with halo mass less massive than this critical value, the 
amplitude of the low mass end of the stellar mass function can be suppressed 
effectively. The functional form in Eq.~\ref{eqn:9} is chosen
in order to fit the amplitude of 
the SMF at the low mass end when applied to both MS and MS-II simulation.

We have also tested a model in which we shut off all SF in the burst mode 
in low mass haloes and keep the SF in the quiescent mode the same as in 
the fiducial model 1). For such a model, the low mass end of the SMF is still 
higher than observation. This indicates that SF in both the quiescent and 
the burst mode must be modified simultaneously to suppress the numbers of 
low mass galaxies effectively. Note that the values 
of the power law indices that determine the dependence of SF on halo mass 
for quiescent and burst modes might have some degeneracy. 
However, we do not study the degeneracy of the two modes of SF in this work 
any further, but focus on the qualitative effect of modifying SF in each mode.

\subsection {Model 3b}
In model 3b), we test the effect of including a dependence of the SF 
efficiency on Hubble time. Model 3b) is almost identical to model 3),
except that the SF efficiency in the quiescent mode is assumed to depend on
cosmic time, in the same way as for the fiducial model 1). 
The SF efficiency is rescaled to match again the
low mass end of SMF in SDSS observation, resulting in:

\begin{equation}
{f_s} = 1.74\,{M_{12}^{0.94}10^{-0.30[\log M_{12}]^2}}t^{-0.82}\,.
\label{eqn:10}
\end{equation}

The SF efficiency in model 3b) is also shown in Fig.~\ref{fig:SFR}. It
is higher than the SF efficiency in model 3) at high redshift, and is lower
at $z=0$. We will see in the next section that compared with model 3),
model 3b) results in a similar SMF at $z=0$, and higher amplitude of SMF 
at higher redshifts of $z\sim 2-3$. The lower SFR at redshift $0$ results 
in low mass galaxies with a lower SSFR than in model 3).
Besides, model 3b) predicts a higher SFR density at redshifts higher than
$\sim 2$, and a lower SFR density at lower redshifts. The general results
are comparable with model 3) though.

\subsection{Model 4}
With a modified SF law in low mass haloes, model 3) and 3b)
are able to reproduce many
observational statistics of low mass galaxies, as we will show in
in Sec.~\ref{sec:results}. However, the properties
of high mass galaxies differ significantly from observations.
The most obvious deviation is that the modeled massive galaxies are in
general too active. This is mainly due to the inclusion of slower
stripping of hot gas for satellite galaxies in our models.
The deviation can be alleviated by allowing for less efficient cooling
in massive haloes. Physically, this corresponds to mechanisms like
stronger AGN feedback effect that prevents gas from cooling.

In model 4), we apply further modifications of cooling and SF to
model 3), focusing on massive galaxies. Model 4) is presented as a simple
test to see if the properties of massive galaxies can be better fitted,
while keeping the the treatment of low mass galaxies unchanged.
The modifications include:

\begin{itemize}
  \item Lower cooling efficiencies are assumed for haloes more massive
than $10^{11.75}h^{-1}{\rm M}_{\odot}$, as shown in Fig.~\ref{fig:coolingmodel4} in
the Appendix.

  \item SF in both the quiescent and the burst mode is stopped completely in
haloes more massive than $5\times10^{12}h^{-1}{\rm M}_{\odot}$ at $z<1.3$.

  \item The dynamical friction time is assumed to be dependent on Hubble time 
and is shorter at higher redshift, with $\alpha_{\rm df}=5\times(t/13.6)^{0.5}$
instead of $\alpha_{\rm df}=5$ in model 3) \citep[see][]{weinmann2011a}.
\end{itemize}

These modifications are done to fit the observed properties of 
massive galaxies better. The first two modifications make massive galaxies 
much more passive than in model 3). The change in the dynamical friction 
time follows the idea of \citet{weinmann2011a}. As discussed there,  
the merger time in the standard model may be overestimated by an order 
of magnitude at high redshift, mainly due to the more radial 
orbits of high redshift satellite galaxies \citep{dekel2009, hopkins2010}.  
Therefore we assume a time-dependent dynamical friction time, which gives 
a better fit to the SFR density and SMF at $z>2$, and does not
affect the other statistics studied in this work much.

\section{Results}
\label{sec:results}

In this section, we show statistical results for
the galaxy population produced by the models presented
in the last section, including galaxy SMF at both low and high redshifts,
the specific SF rate--stellar mass relation, the cold gas mass function, the
projected two point correlation functions, and SF rate density as a function
of redshift. We compare these results with observations, and analyze the
effect of different modifications of SF laws on those statistics.

\subsection {Stellar mass function at $z=0$}
\begin{figure}
\bc
\hspace{-0.4cm}
\resizebox{8cm}{!}{\includegraphics{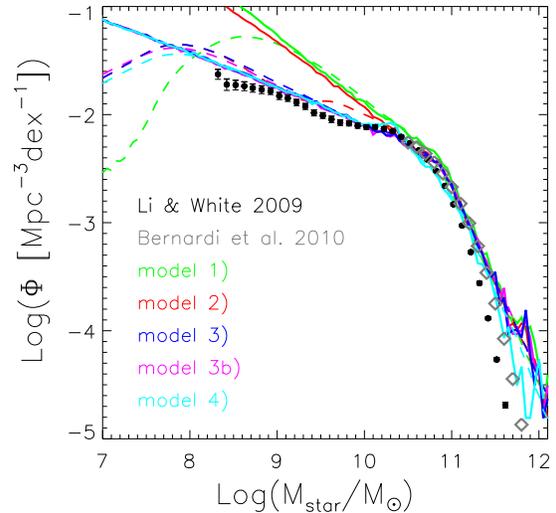}}\\%
\caption{Stellar mass functions at $z= 0$. Green, red, blue, magenta and cyan
lines are for model 1), 2), 3), 3b) and 4) respectively. Solid and 
dashed lines are for MS-II and MS results. Black symbols with error bars 
are based on SDSS DR7 including both
counting and cosmic variance uncertainties \citep{li2009, guo2010a}. 
The grey diamonds show the functional fit 
to the SMF from \citet{bernardi2010},
for galaxies more massive than $10^{10.5}{\rm M}_{\odot}$.
}
\label{fig:SMF}
\ec
\end{figure}

The SMF at $z=0$ is the main quantity that was used to 
constrain the parameter values in our models, and is plotted in 
Fig.~\ref{fig:SMF}. 
Similar to previous SAMs \citep[DLB07,][]{guo2010}, the fiducial model 1) 
predicts too many low mass galaxies. 
As a result of the modifications in the SF law
in low mass haloes, model 2) applied to the MS 
fits the SMF for stellar masses less massive than $10^{9.3}{\rm M}_{\odot}$. 
However, when applied to the MS-II, model 2) still predicts too many low 
mass galaxies. By suppressing the SF in the burst 
mode in low mass haloes, model 3) 
gives a reasonable fit to the observed SMF, for both MS and MS-II simulations. 
Including a dependence of the SF on cosmic time in model 3b) does not 
affect the SMF noticebly. For model 4) which changes further the cooling 
and SF in high mass haloes, the SMF at z=0 is somewhat lower at the 
massive end, and is more consistent with observations.

As mentioned in sec.~\ref{sec:model}, we have tested that when shutting off
all SF in either quiescent mode or burst mode in haloes less massive than
$10^{11.5}h^{-1}{\rm M}_{\odot}$, the predicted SMF still exceeds observations at 
low mass end, since SF from the other mode compensates. This indicates that SF
in both quiescent and burst mode must be modified simultaneously to fit
the observed SMF, as is done in model 3) and 3b).
This also explains why \citet{lagos2010} find no obvious change in the
resulting SMF when only modifying the SF in the quiescent mode in their SAM.

\subsection {SSFR--$M_{\rm star}$ relation}

\begin{figure*}
\bc
\hspace{-1.4cm}
\resizebox{16cm}{!}{\includegraphics{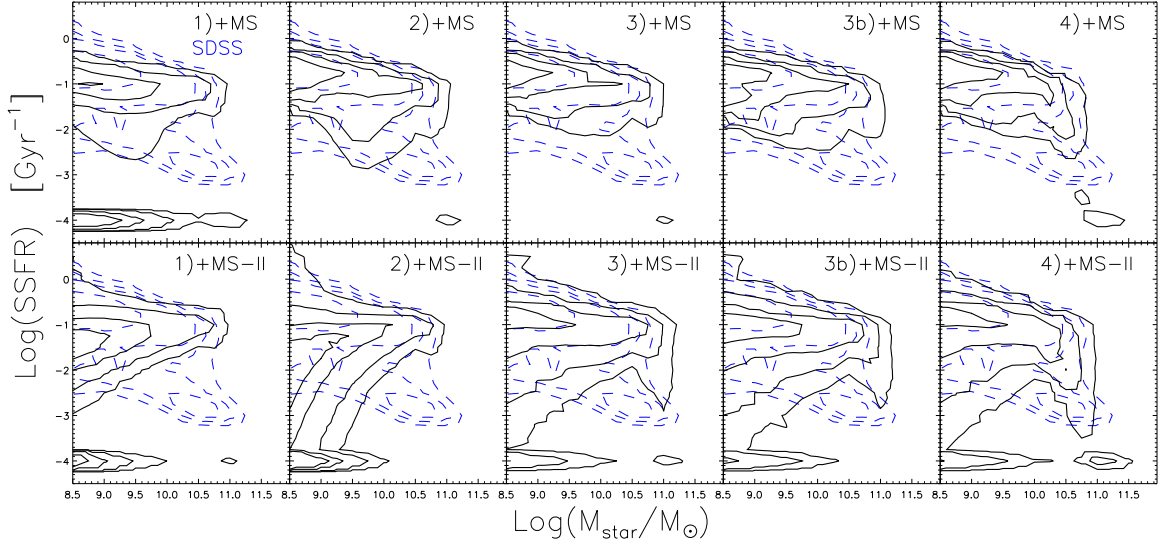}}\\%
\caption{Contour plot of SSFR--$M_{\rm star}$ relation in different
models compared with observation. Upper panels are for models applied to the MS
and lower panels are for MS-II. Blue dashed contours are SDSS
DR7 observation, derived from the MPA-JHU release of the SDSS DR7
catalogue, with stellar masses based on fits to the photometry following
the philosophy of  \citet{kauffmann2003} and  \citet{salim2007}, and SF
rates based on  \citet{brinchmann2004}.
Black contours are the results of
models 1), 2), 3), 3b) and 4). 
When plotting the model results, we set the minimum SSFR 
value to be $10^{-4}${\rm Gyr}$^{-1}$. Contours levels show regions
that enclose 38, 68, 87 and 95 percent of galaxies.
}
\label{fig:SSFR}
\ec
\end{figure*}

The specific SF rate (SSFR) is defined as the
ratio between the SF rate 
and the stellar mass of a galaxy. The relation between SSFR and
galaxy stellar mass is a fundamental observable which needs to be reproduced
by a successful model. Observationally, there is a clear trend that high mass
galaxies are passively evolving with low SSFR, while low mass galaxies have 
high values of SSFR \citep{salim2007, schiminovich2007}. SAMs, however, usually
predict similar, if not lower SSFR in low mass galaxies than that in
massive ones \citep{fontanot2009}. Comparisons that focus on galaxy colours
also show this discrepancy, with low mass galaxies having redder colours 
than observed \citep[Guo et al. 2011, but see also][]{weinmann2011b}. 

\begin{figure}
\bc
\hspace{-0.4cm}
\resizebox{8cm}{!}{\includegraphics{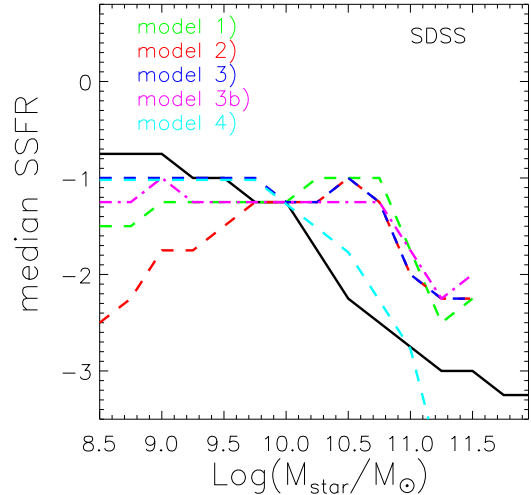}}\\%
\caption{
The median SSFR as a function of galaxy stellar mass. The black solid line 
shows observational results from the SDSS DR7 galaxy sample. Green, red, blue, 
magenta, and cyan lines are predictions from our models 1), 
2), 3), 3b) and 4), combined with MS-II.
}
\label{fig:redfraction}
\ec
\end{figure}

Fig.~\ref{fig:SSFR} shows the SSFR--$M_{\rm star}$ relation in our models,
compared with SDSS observational result. The upper panels are for models
applied to the MS simulation, and the lower panels show the MS-II results.
In each panel, black contours are the model results, while blue contours
are SDSS results. 
Observationally, galaxies reside in two distinct sequences in the SSFR-stellar
mass plot: an active sequence with high SSFR that is more prominent for low mass
galaxies, and a passive sequence with lower SSFR that contains mainly massive
galaxies. The observational location of the passive sequence, however, is
not well determined, due to the uncertainty on measuring the SSFR for galaxies
with little SF \citep{salim2007}. 
In addition, Fig.~\ref{fig:redfraction} shows the median SSFR as a 
function of galaxy stellar mass. Colored lines are predictions from our 
models combined with MS-II. The black solid line is the result of SDSS DR7 
galaxy sample, which shows clearly that more massive galaxies have 
in general lower SSFR.

Comparing the model results applied to the MS and MS-II simulations
respectively in Fig.4, it is obvious that the resolution of simulation has  
a large effect on the modeled SSFR for low  mass  galaxies,
especially for models 1) and 2). This highlights the fact that resolution 
can have a large impact on semi-analytical model predictions (see also 
Guo et al. 2011). We note that the resolution dependency is weaker for 
models 3)-4), only affecting the passive sequence of galaxies whose 
location is anyway uncertain (see above).
To compare the model prediction with observation for low mass galaxies, 
we focus on results of the models combined with the MS-II simulation.
Similar to previous SAMs, model 1) predicts more passive low mass galaxies than
observed, and the active sequence in model 1) is lower than observed. 
Its predicted median SSFR is lower than the SDSS result for 
galaxies less massive than $10^{9.7}{\rm M}_{\odot}$. When modifying the SF in the
quiescent mode in low mass haloes in model 2), more low mass galaxies become 
passive, and the median SSFR is even lower than in model 1). This is due to 
the in general lower SF efficiency assumed in model 2)
for haloes less massive than $\sim 10^{11.5}h^{-1}{\rm M}_{\odot}$, as shown in 
Fig.~\ref{fig:SFR}.

Suppressing the SF in the burst mode in low mass haloes, model 3) and 3b) 
predict a lower fraction of passive galaxies, with higher median SSFR for 
low mass galaxies. This is because the lower burst efficiency leaves 
galaxies with more cold gas by $z=0$. The results are then in 
reasonable agreement with observations.
When comparing these two models in detail, model 3b) predicts on average 
more passive low mass galaxies and is therefore a less good match to
observations. The location of the active sequence is lower than in
observation, and thus not as well reproduced as in model 3).

For all models 1)-3b), the SSFR for massive galaxies are significantly 
overpredicted by the model. This is corrected in
model 4), by modifying the
cooling and SF in massive haloes. Overall, model 4) is in good agreement with 
the observed SSFR-$M_{\rm star}$ relation at all stellar masses.

\subsection {Cold gas mass function}

\begin{figure}
\bc
\hspace{-0.4cm}
\resizebox{8cm}{!}{\includegraphics{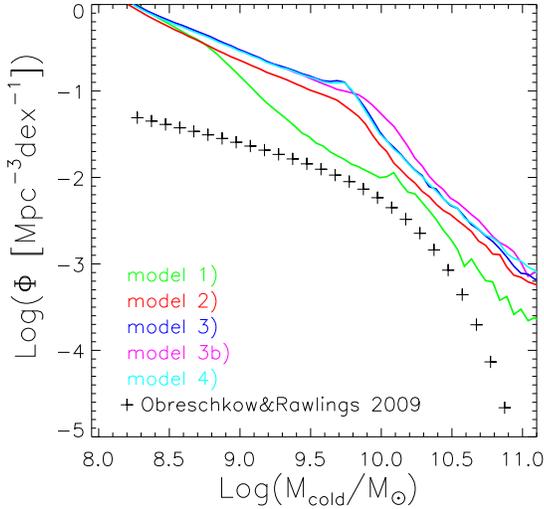}}\\%
\caption{Cold gas mass functions at z=0.  Green, red, blue,
magenta and cyan lines are for model 1), 2), 3), 3b) and 4), for MS-II results.
Cross symbols show the Schechter function fit for the cold gas mass function
according to \citet{obreschkow2009}.
}
\label{fig:CMF}
\ec
\end{figure}

Fig.~\ref{fig:CMF} presents the cold gas mass functions at z=0. Green, red, 
blue, magenta and cyan lines are for model 1), 2), 3), 3b) and 4), 
using the MS-II results. Crosses show the Schechter 
function fit to the cold gas (HI+H$_2$+helium) mass function 
given in \citet{obreschkow2009}, which is derived from the HI mass function 
of HIPASS observation \citep{zwaan2005}, combined with a theoretical model to 
describe the H$_2$/HI mass ratio. As pointed out in 
\citet{obreschkowCroton2009}, semi-analytic models only distinguish between 
the hot and the cold gas phase. However, observations have revealed the 
existence of a significant amount of warm and ionized gas in the Milky 
Way \citep{reynolds2004}. This motivated \citet{obreschkowCroton2009} 
to devide their model cold gas mass function by a factor of $1.45$. We 
follow their example here, but we stress that this factor is obviously 
uncertain and may well be higher, depending on galaxy mass and redshift.

The cold gas mass function of the fiducial model 1) exceeds 
observations. As expected, it is somewhat higher than 
for the DLB07 model \citep{obreschkowCroton2009, fu2010}, because of
the slower stripping of hot gas for satellites. The other models predict 
even higher amplitudes. With on 
average lower SF efficiencies in low mass haloes in model 2), 3) and 3b), 
there is much more cold gas left in the galaxies. This is not surprising, 
since we have left cooling and SN feedback efficiencies at the default
DLB07 values. Model 4) predicts similar result of cold gas mass function
as in model 3).

Although there are some uncertainties related to the determination of cold
gas mass function in observation \citep{keres2003,obreschkow2009, fu2010},
it seems that our models with modified SF  predict drastically too much
cold gas mass in galaxies. Thus, while the modifications 
to the SF law we suggest here improve 
the agreement between the statistical properties of the galaxy population 
in many aspects, our simple model is
clearly not the final answer to all the problems that exist.
This will be discussed in more detail in section 4.

\subsection{Correlation function}

The two point auto-correlation function of galaxies is a fundamental
measure of the spatial distribution of a certain population of galaxies 
and is well determined by observations \citep[e.g.][]{li2006}. It is however 
not well fitted by current SAMs. For example,  \citet{guo2010} over-predict 
the correlation functions for low mass galaxies and on small scales.
They suggest that the over-prediction of clustering of galaxies on small 
scales in the models can be explained by the too high value of $\sigma_8$
adopted by the Millennium simulation, which is $0.9$ compared to $0.81$
suggested by the WMAP-7 year result \citep{komatsu2011}. However,
it is not certain that the Guo et al. SAM combined 
with the correct cosmology would give correlation functions in agreement 
with observations \citep[see][]{wangjie2008}. Here
we test the effect of modifying the SF law
on the resulting correlation functions using the same underlying 
$N$-body simulations as \citet{guo2010}.

\begin{figure*}
\bc
\hspace{-1.4cm}
\resizebox{16cm}{!}{\includegraphics{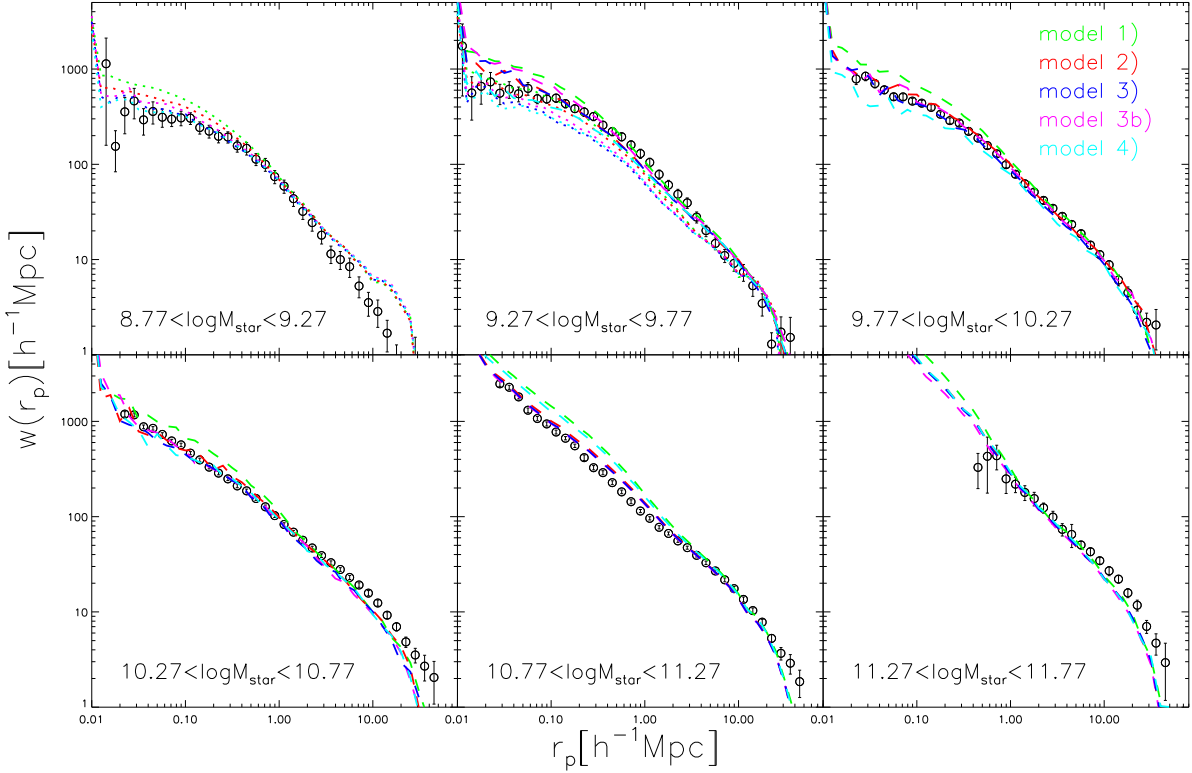}}\\%
\caption{Projected two point auto-correlation functions of galaxies in
different stellar mass bins. Circles with error bars are SDSS
DR7 results \citep{guo2010}, calculated with the same method as 
presented in \citet{li2006}.
Green, red, blue, magenta, and cyan lines are for 
model 1), 2), 3), 3b), and 4) respectively. Dotted (dashed) lines are for 
models based on the MS-II (MS) results. For stellar mass bins with 
$\log(M_{\rm star})$
$>$ $9.77$, only MS results are shown. For the lowest mass bin,
only MS-II results are presented.
}
\label{fig:CF}
\ec
\end{figure*}

In our models,
galaxy positions are determined by the positions of haloes/subhaloes they
reside in. For galaxies that have lost their host subhalo due to stripping, 
we use the location of the most-bound-particle of the
last identified subhalo. Since we use the same dynamical friction 
prefactor of $\alpha_{\rm df}=5$ for models 1), 2), 3), 3b) and 4), galaxy 
locations in all these models are exactly the same. The only differences are
in the stellar masses of galaxies that vary due to the different SF laws used. 
Also, the number of galaxies might be slightly different due to the effect of 
the stellar mass on the dynamical friction time. 
Consequently, the differences in the correlation 
functions between the models are mainly
due to the different stellar mass assigned to each galaxy.

Fig.~\ref{fig:CF} shows the projected correlation functions of galaxies in
different stellar mass bins, computed in the same way as in 
\citet{neistein2011}. 
Model 1) over-predicts the correlation functions for galaxies less 
massive than $10^{11.27}{\rm M}_{\odot}$, and at scales smaller 
than $\sim 1h^{-1}$Mpc. With a modified SF law in low mass haloes, 
correlation functions become lower 
at small scales for low mass galaxies, which brings model 3) into agreement 
with the observational results at all stellar masses. 
The success of model 3) in reproducing the correlation functions  
shows clearly that changes in the  
SF law, that lead to changes in the relation between stellar 
mass and subhalo mass, can significantly affect correlation functions. Thus, 
it is in general very hard to say whether a mismatch with observed correlation
functions indicates a problem with baryonic recipes, or with cosmology. We thus 
confirm the results by  \citet{wangjie2008} who find that a similar
match to observed correlation functions can be obtained by SAMs using different 
cosmological parameters, depending on the detailed baryonic recipes.

Model 3b) gives similar results to model 3) for galaxies more massive than
$10^{10.27}{\rm M}_{\odot}$. For lower mass galaxies, the correlation functions 
of model 3b) are 
higher than for model 3) on small scales. Correlation functions of galaxies 
in model 4) are in general similar to model 3), with a somewhat worse fit 
in the stellar mass bin of $\log M_{\rm star}$=[10.77, 11.27].

\subsection{Stellar mass functions at high $z$}

At redshifts less than $\sim 0.8$, current SAMs like \citet{monaco2007} 
and \citet{somerville2008} predict SMF consistent with observations. 
At higher redshifts, however, models normally over-predict the abundance of 
galaxies less massive than $10^{10}{\rm M}_{\odot}$ \citep{fontanot2009,guo2010}. 
The same is true for the K-band luminosity function \citep{henriques2011}.
This may indicate that SF at redshifts above 0.8 is not modeled correctly
in these models.

\begin{figure*}
\bc
\hspace{-1.4cm}
\resizebox{12cm}{!}{\includegraphics{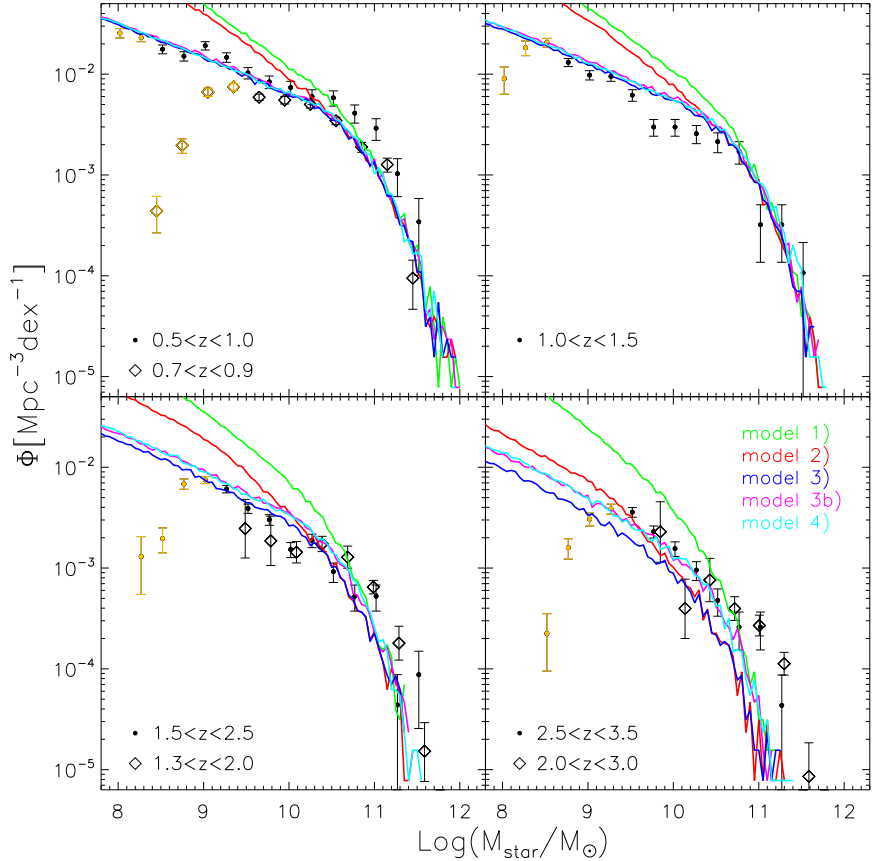}}\\%
\caption{Stellar mass functions at redshifts above zero. Green, red, blue,
magenta and cyan lines are for model 1), 2), 3), 3b) and 4). Model results
combined with MS-II are shown at redshifts of $0.8, 1.2, 2.0, 3.0$.
Black points are observational results from  \citet{kajisawa2009}, for all 
four redshift bins. In the upper left panel, observational results from 
\citet{pozzetti2007} at $0.7<z<0.9$ are plotted as diamonds. Gold symbols 
are the data points below the completeness limit
in these two observational samples. In the lower two panels, observations from
 \citet{marchesini2008} for $1.3<z<2.0$ and $2.0<z<3.0$ are shown as diamonds. 
Observed stellar masses of galaxies are normalized to the Chabrier 
IMF \citep{chabrier2003}. 
}
\label{fig:highzSMF}
\ec
\end{figure*}

Fig.~\ref{fig:highzSMF} gives the results of SMF at higher redshifts in
different models studied in this work. Model results are presented at redshifts
of $0.8$, $1.2$, $2.0$, $3.0$, and are convolved with a Gaussian error of 
deviation $0.25$ dex in $\log M_{\rm star}$, to account for the various errors 
in estimating stellar masses in observations  \citep{fontanot2009}. 
Observation results are shown at comparable redshift ranges to the models. 
Gold symbols indicate data points below the limiting stellar mass of the 
observed galaxy samples, where incompleteness could be significant 
\citep{pozzetti2007,kajisawa2009}. All observational stellar masses of
galaxies are normalized to the Chabrier IMF  \citep{chabrier2003}, to be 
consistent with the previously shown SDSS SMF, and the model derived SMF.

The fiducial model 1) predicts too many low mass galaxies at all redshifts.
With a modified SF law in both modes, model 3) and 3b) both give consistent 
SMF with observations up to redshifts of around $3$. Differences of these 
two models can be seen at redshifts higher than $2$, where model 3b) 
predicts a SMF with a higher amplitude. 
This is because the SF efficiency in model 3b) depends on 
time and is higher at high redshifts.

In model 4), with the assumption that the dynamical friction time 
is shorter at high redshifts, SMF at redshifts of around $2-3$ are higher
than in model 3), and somewhat  closer to observations. The SMF in model 4) 
happen to be quite similar to the results of model 3b).
This reflects the degeneracies inherent in SAMs, in this case
between the dependence of dynamical friction time on redshift, and the
dependence of SF rate on redshift.

\subsection{SFR density}

\begin{figure}
\bc
\hspace{-0.4cm}
\resizebox{8cm}{!}{\includegraphics{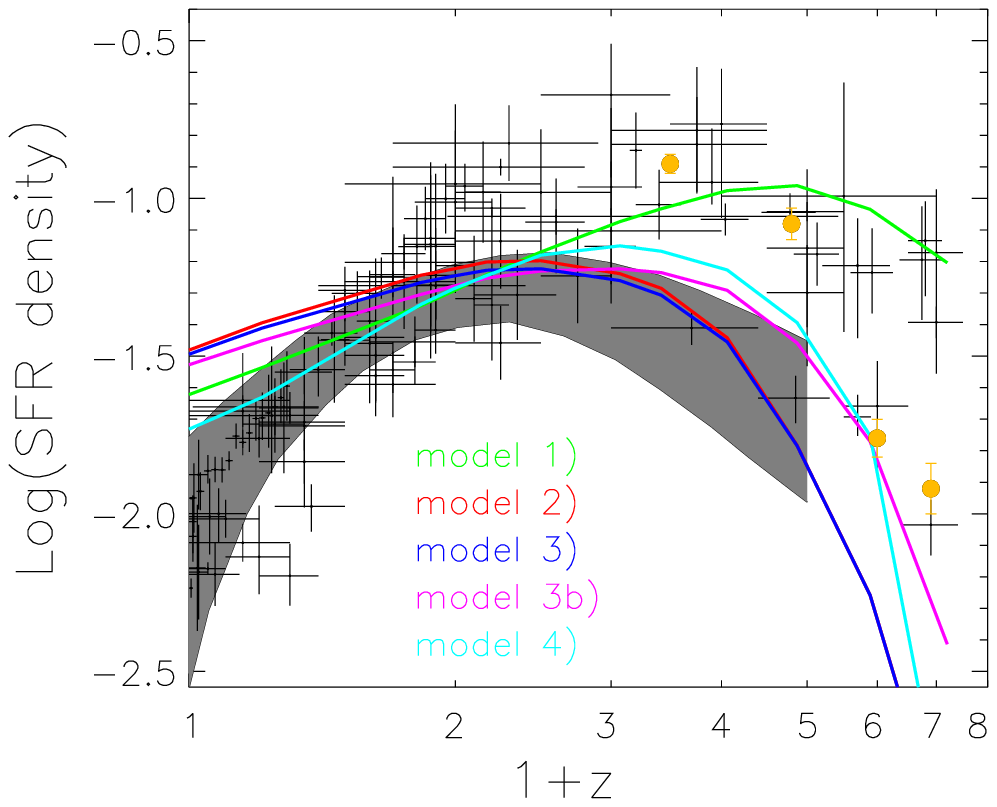}}\\%
\caption{Cosmic star formation density as a function of redshift.
Green, red, blue, magenta and cyan lines are for
models 1), 2), 3), 3b) and 4) respectively, based on the MS simulation.
Black crosses are observational estimates compiled by  \citet{hopkins2007}.
The grey shaded region shows the 1-$\sigma$ confidence level of
the observational result, as compiled by  \citet{wilkins2008}. Gold symbols 
are the results of  \citet{bouwens2009}, including the contributions
from highly dust obscured galaxies and ULIRGs. 
Stellar masses are normalized to the Chabrier IMF \citep{chabrier2003}.
}
\label{fig:SFRdensity}
\ec
\end{figure}

Fig.~\ref{fig:SFRdensity} shows the star formation rate (SFR) density as a
function of redshift, as predicted by different models combined with MS 
simulation. Black crosses are observational results compiled by  
\citet{hopkins2007}. The grey shaded region shows the 1-$\sigma$ confidence 
level of the observational results by  \citet{wilkins2008}, which 
are derived indirectly from the evolution of the stellar mass function. 
Gold symbols are the results of \citet{bouwens2009}, including the 
contributions from highly dust obscured galaxies and ULIRGs.
Stellar masses are normalized to the Chabrier IMF \citep{chabrier2003}.

The SFR density of the fiducial model 1) shows a continuous increase with 
redshift and peaks at a redshift of around $3-4$, which is 
clearly a higher redshift than in observations.
This offset is similar to the one present in the SAM
of Guo et al. (2011). With a modified SF efficiency
in both the quiescent and the burst mode, the SFR density of model 3) and 3b)
drops dramatically at higher redshifts than $z\sim2$.
Due to the dependence of SF on Hubble time assumed in model 3b),
this model gives a higher SFR density at high $z$, and a lower SFR density 
at low $z$ than model 3). At low redshifts, both model 3) and 3b) predict 
SFR density higher than the fiducial model 1), which lies slightly 
above the observational values. 

With further suppression of cooling and SF in massive haloes in model 4),
the observed sharp decline of SFR density towards low redshifts appears. 
Predictions of model 4) are within the observational constraints, 
while the SFR density peaks at redshift of around $2$. 
Recently, \citet{magnelli2011} studied
the evolution of dusty infrared luminosities function using Spitzer data.
Assuming a constant conversion between the IR luminosity and SFR, they find
that the SFR density of the Universe strongly increases towards $z=1.3$,
and stays constant out to $z=2.3$. Model 4) matches the
result of their observation.

\section{Discussions and Conclusions}
\label{sec:conclusion}

We use the method developed by  \citet{NW2010}, combined with 
both the Millennium (MS) and Millennium-II (MS-II) cosmological 
simulations, to study the effect of modifying the star formation (SF) 
recipe in low mass galaxies. We show that by modifying SF in both the 
quiescent and the burst mode, the stellar mass function 
observed in the local Universe
can be reproduced well down to $10^{8.5}{\rm M}_{\odot}$. Simultaneously, 
the models can fit the observed median SSFR--$M_{\rm star}$ relation 
for galaxies less
massive than $10^{10}{\rm M}_{\odot}$, the correlation functions for galaxies 
more massive than $10^{8.77}{\rm M}_{\odot}$, 
the stellar mass functions up 
to redshift of around $3$, and the general trend of the SFR density as a 
function of redshift. 

The modifications to the SF recipe in our models with respect to 
standard SAMs (e.g. DLB07) include:
\begin{enumerate}
\item no sharp threshold in the cold gas mass for SF; 
\item letting the SF efficiency in the quiescent mode depend on host halo mass;
\item removing the dependence of the star formation rate on Hubble time 
(which came via the dependence on disk dynamical time);
\item a lower star burst efficiency in low mass haloes;
\item additional modifications of cooling and SF in massive haloes to match
the properties of high mass galaxies.
\end{enumerate}

Model 2) only includes the changes to the quiescent mode of
SF, i) - iii). We show that this is not enough to reproduce the low
mass end of the SMF, as star formation in the burst mode compensates
for the changes to the SF law in 
the quiescent mode. This is also the reason why \citet{lagos2010}
have found that their changes to the SF law in the quiescent
mode does not change the resulting SMF much. In model 3), we have
thus additionally decreased SF in the burst model (modification iv), 
which results in a clearly improved SMF. We also considered a model 3b) which 
is similar to model 3), except that we allow for the usual time-dependence
of the SF law (i.e. do not make modification iii). Results of model 3b) are
similar to model 3) out to $z \sim 2$, except that the SSFR--$M_{\rm star}$ 
relation is slightly less well reproduced. It is thus not clear whether 
modification iii) is necessary.

Note that removing the time-dependence of the star formation efficiency 
is a significant change with respect to previous 
models. Even in the recent models of \citet{fu2010} 
and \citet{lagos2010,lagos2011}, where SF efficiency does not depend
on the disk dynamical time of galaxies, a time-dependence enters 
via the conversion efficiency from atomic to molecular gas, that depends 
on the gas density. In order to justify the behaviour we suggest in 
model 3), we would need to postulate a mechanism that scales with time in 
the opposite way than usually assumed, like for example a metallicity-dependent
conversion of atomic to molecular gas \citep{krumholz2011}.
We note that a weak dependence of the SF efficiency on cosmic time 
was seen in hydrodynamical simulations \citep{neistein2011c}, although
the reason for this behaviour is still not clear.

In our models 2) -- 4), the ratio between the quiescent star formation rate 
and the cold gas mass, $\dot{M}_{\rm star}/M_{\rm cold}$ which is equal to 
the gas consumption timescale, is roughly proportional to $M_{\rm halo}^{2}$ 
for low mass haloes and is almost independent of halo mass 
for high masses. This may in fact be supported by recent observations. 
Recently, \citet{shi2011} derive an extended Schimdt law from an
observed galaxy sample that extends over 5 orders of magnitude in 
stellar density, including galaxies with low surface brightness. 
They find that $\dot{M}_{\rm star}/M_{\rm cold}$ is proportional to
$M_{\rm star}^{0.52}$, with a 1-$\sigma$ scatter of $0.4$ dex. 
The stellar mass of galaxies has been claimed to obey a tight relation
with the host halo mass \citep{conroy2009, moster2009, guo2010a}. 
It is proportional to $M_{\rm halo}^{2.8}$ for low mass
haloes and to $M_{\rm halo}^{0.2}$ at high mass end \citep{ wang2006}. Without
considering the scatter of the relation, this indicates that the observed 
$\dot{M}_{\rm star}/M_{\rm cold}$ is roughly proportional to $M_{\rm halo}^{1.46}$
at low masses and $M_{\rm halo}^{0.1}$ at high mass end. The dependence
on halo mass in our models is therefore quite close to the observational
result by  \citet{shi2011}.

In all our models, the cold gas mass function of galaxies is 
dramatically over-predicted. This is because in DLB07, the total amount of 
cold gas and stellar mass exceeds the observed total amount 
\citep[see][]{obreschkowCroton2009, fu2010}.
Decreasing SF rates in low mass galaxies while letting 
cooling and feedback recipes remain unchanged naturally results in an 
overproduction of the cold gas mass in low mass galaxies. This is not 
only a problem for the models we present here. As shown recently 
by \citet{lu2011}, when the model K-band luminosity function is 
forced to fit the data in the local Universe, the cold gas 
mass function is dramatically over-predicted in all the semi-analytic models 
they study. This is consistent with our models over-predicting the 
cold gas mass functions when we fit the stellar mass function at z=0.
On the other hand, with similar cooling and supernovae feedback 
recipes, the models that do fit the cold gas mass function in turn have 
problems in reproducing the stellar
mass functions \citep[e.g.][]{fu2010,obreschkowCroton2009}. 
In the meantime, approaches like increasing the SN feedback, 
that suppress the total amount of cold gas and stellar mass, 
lead to several other serious problems \citep{guo2010}.
These results show again that it is currently difficult for a single model 
to fit all observations, as mentioned in Sec.~\ref{sec:intro}, 
unless we allow for free tuning of all recipes, including cooling 
(see NW2010).

Although the modifications of the SF law presented in this work 
help to improve the agreement with several observed statistical properties 
of galaxies, and seem to follow a similar scaling like the observed SF 
efficiencies in galaxies, the normalization of the SF efficiency cannot be
correct. This is clear from the fact that our models overpredict
the cold gas mass function. With a lower cold gas fraction in galaxies, 
the SF efficiency would obviously have to
be higher than currently assumed in the models, to obtain the same SF rate. 

\begin{figure*}
\bc
\hspace{-1.4cm}
\resizebox{14cm}{!}{\includegraphics{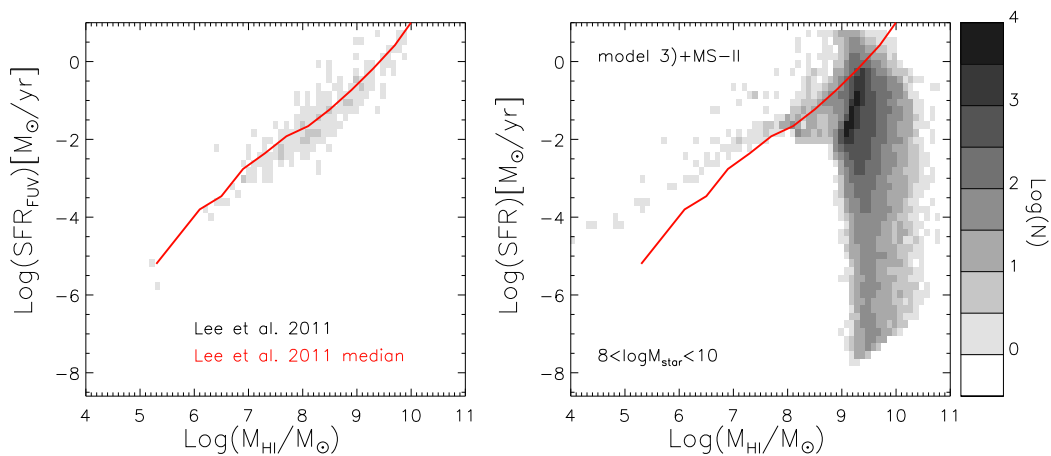}}\\%
\caption{SF rate -- HI gas relation in model 3) compared with 
observation. The left panel is the FUV SFR--HI mass results of \citet{lee2011}.
The right panel shows results of 3) combined with
MS-II simulations, for galaxies with $8<\log(M_{\rm star})<10$. Gray scales 
correspond to the logarithm of the number of galaxies. Red line in each
panel is the median relation of the observed galaxies of  \citet{lee2011}.
}
\label{fig:SFlaw}
\ec
\end{figure*}

In Fig.~\ref{fig:SFlaw}, we compare the SFR--HI mass 
relation in model 3) combined with MS-II (right panel), with a recent 
observation of the FUV derived SFR--HI mass relation for galaxies within 
$\sim 11$ Mpc of the Milky Way \citep[][left panel]{lee2011}. The cold 
gas mass in model 3) is converted to the HI mass to be 
compared with observation, using a correction including three factors. First,
as for Fig.~\ref{fig:CMF}, the cold gas mass from model 3) is divided
by a factor of $1.45$, to account for a warm ionized gas phase, as done in
\citet{obreschkowCroton2009}. Second, we multiply the results by 
a factor of $0.76$  to remove the contribution of helium and heavier
elements  \citep{power2010}. Finally, the hydrogen gas mass is divided by $1.4$,
as adopted by \citet{power2010} and \citet{lu2011}, to remove the contribution 
of $H_2$. Since the galaxies in the observed sample have stellar masses 
less than $\sim 10^{10}{\rm M}_{\odot}$,
we present model results for galaxies with $8<\log (M_{\rm star})<10$. 
The median relation from the observations is plotted as red line in 
each panel. Although the observations of \citet{lee2011} are limited to 
a small volume of space, the obviously different relations in observations
and in model 3) indicate that the SF efficiency 
in model 3) is indeed much lower than in reality. 

Fitting the most important observed properties of galaxies is not a trivial 
task. It is not clear 
up to now if this difficulty in modeling galaxy properties reflects a 
fundamental problem in our understanding of the dark matter universe,
or if it is mainly due to an insufficient understanding
of the baryonic physics involved in galaxy formation and evolution.
For example, perhaps SAMs miss an important ingredient of galaxy formation, 
like a mechanism that preheats the gas in the universe so that
it cannot cool to low mass haloes 
\citep[][but see also \citet{crain2007}]{mo2005}, or a form of feedback
that mainly heats low entropy gas in high redshift haloes \citep{McCarthy2011}.
Alternatively, it could be that cooling is over-efficient in
the current SAMs for some reason. 

With the tests carried out in this work, we show that only modifying 
star formation in the SAMs can already improve agreement with 
observations in several important aspects. Up to now,
a high-resolution SAM that matches the SSFR-stellar mass
relation at $z=0$, the SMF and its evolution, the correlation functions and the 
cold gas fractions of galaxies simultaneously, does not yet seem to exist.
To find such a model, approaches that allow scanning of a large parameter space
\citep{henriques2009,lu2010,bower2010}, and approaches that 
allow deviations from the usually assumed functional forms 
for physical recipes \citep{NW2010}, or that include other physical 
processes than currently considered \citep{henriques2010} may be promising. 
However, even if one or several models are found that do indeed
reproduce all these fundamental observables, 
it will be important to identify degeneracies, and to verify whether the models
are physically plausible and can be brought into agreement with alternative
approaches, like predictions from hydrodynamical simulations.

\section*{Acknowledgments}

We thank the referee for a constructive report to help to improve the 
manuscript.
We acknowledge Cheng Li for providing the SDSS data results and for helpful
discussions, Janice C. Lee for providing the data values in their paper,
and Jian Fu for helpful discussions.
LW acknowledges support from the National basic research program
of China (973 program under grant No. 2009CB24901), the NSFC grants program 
(No. 11143006, No. 11103033, No. 11133003), the Young Researcher Grant of 
National Astronomical Observatories, Chinese Academy of Sciences,
and the Partner Group program of the Max Planck Society.
The Millennium Simulation and the Millennium-II Simulation were
carried out as part of the programme of the Virgo Consortium on the
Regatta and VIP supercomputers at the Computing Centre of the
Max-Planck Society in Garching.
The halo/subhalo merger trees for the Millennium
and Millennium-II Simulations are publicly available
at http://www.mpa-garching.mpg.de/millennium

\bsp
\label{lastpage}

\bibliographystyle{mn2e}
\bibliography{SF}


\appendix

\section{Cooling efficiencies }

\begin{figure}
\bc
\hspace{-0.4cm}
\resizebox{8cm}{!}{\includegraphics{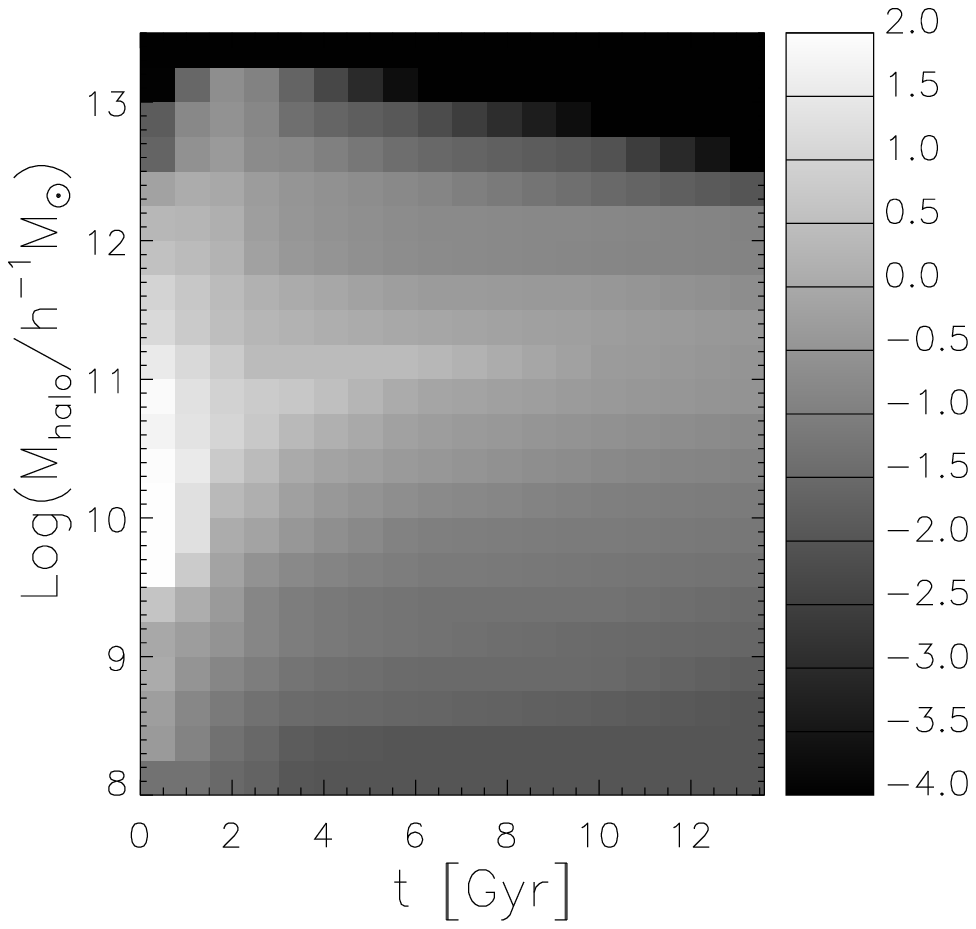}}\\%
\caption{Cooling efficiencies as a function of halo mass and cosmic time in
model 0), the fiducial model 1), and also in model 2) and 3), 3b). The
gray scale shows Log values of the cooling efficiencies in units of 
log$[{\rm Gyr}^{-1}]$.
X-axis is the cosmic time, with $13.7$ Gyr corresponding to the present day.
}
\label{fig:cooling}
\ec
\end{figure}

Our model 0) is similar to model 0 of NW2010, with a few modifications
to the cooling efficiencies, which we explain below. Model 0 of NW2010 is 
adapted to the MS simulation, and thus only 
includes efficiencies down to the resolution limit of this simulation, 
which is $\sim 10^{10}h^{-1}{\rm M}_{\odot}$ in halo mass. To apply 
it to the MS-II simulation, we need efficiencies down to a lower halo mass of 
$\sim 10^{8}h^{-1}h^{-1}{\rm M}_{\odot}$. This is straightforward for most 
processes, as they are parametrized by functional forms which can easily 
extended to lower halo masses. The only exception are the cooling 
efficiencies, $f_c$, defined as $\Delta m_{\rm cool}= f_c m_{\rm hot} \Delta t$, 
where $\Delta m_{\rm cool}$ is the amount of gas that is cooled
within a time-step $\Delta t$, and $m_{\rm hot}$ is the mass of hot gas. 

In DLB07, the treatment of gas cooling follows the description 
of \citet{croton2006}, where cooling efficiencies are calculated
according to \citet{white1991}, assuming an isothermal gas density profile. 
In model 0 of NW2010, cooling efficiencies are median values
computed from a large statistical sample of galaxies in DLB07, for each
bin of halo mass and cosmic time. The tabulated values as a function
of halo mass and time in NW2010 follow no specific functional 
form and can thus not easily be extended to lower mass haloes.
The obvious solution would be to extract those values from the DLB07 SAM
as applied to MS-II, but this is not possible for technical reasons. Therefore, 
we estimate cooling efficiencies for low mass haloes such that (i)
the general trends of cooling efficiency as a function of halo mass
and redshift are preserved, and (ii) the resulting low mass end slope and
amplitude of stellar mass function at z=0 is similar to the result of DLB07
when applied to MS-II Simulation (see Fig. \ref{fig:fiducial}). In this way 
our estimates of the cooling efficiencies at a given halo mass
should be similar as in the DLB07 model, when averaged over all
redshifts. However, we note that the cooling efficiencies
in a given redshift and halo mass bin may differ.
Apart from extending the cooling efficiencies to lower mass haloes, we
also apply some smoothing to the original cooling
efficiencies found by NW2010, in order to smoothen the stellar mass
functions. 

Fig.~\ref{fig:cooling} shows cooling efficiencies as a function of
halo mass and cosmic time in model 0). These values are given explicitly
in Table \ref{tab:cooling0}. The extrapolation and smoothing to the cooling 
efficiencies used in model 0 of NW2010 can be seen by comparing 
Fig.~\ref{fig:cooling} with Fig. 6 in NW2010, and also comparing the 
values listed in Table 1 with Table 6 of NW2010. The cooling efficiencies 
shown in Fig.~\ref{fig:cooling} and Table \ref{tab:cooling0} are also 
applied to model 1), 2), 3) and 3b).

\begin{figure}
\bc
\hspace{-0.cm}
\resizebox{8cm}{!}{\includegraphics{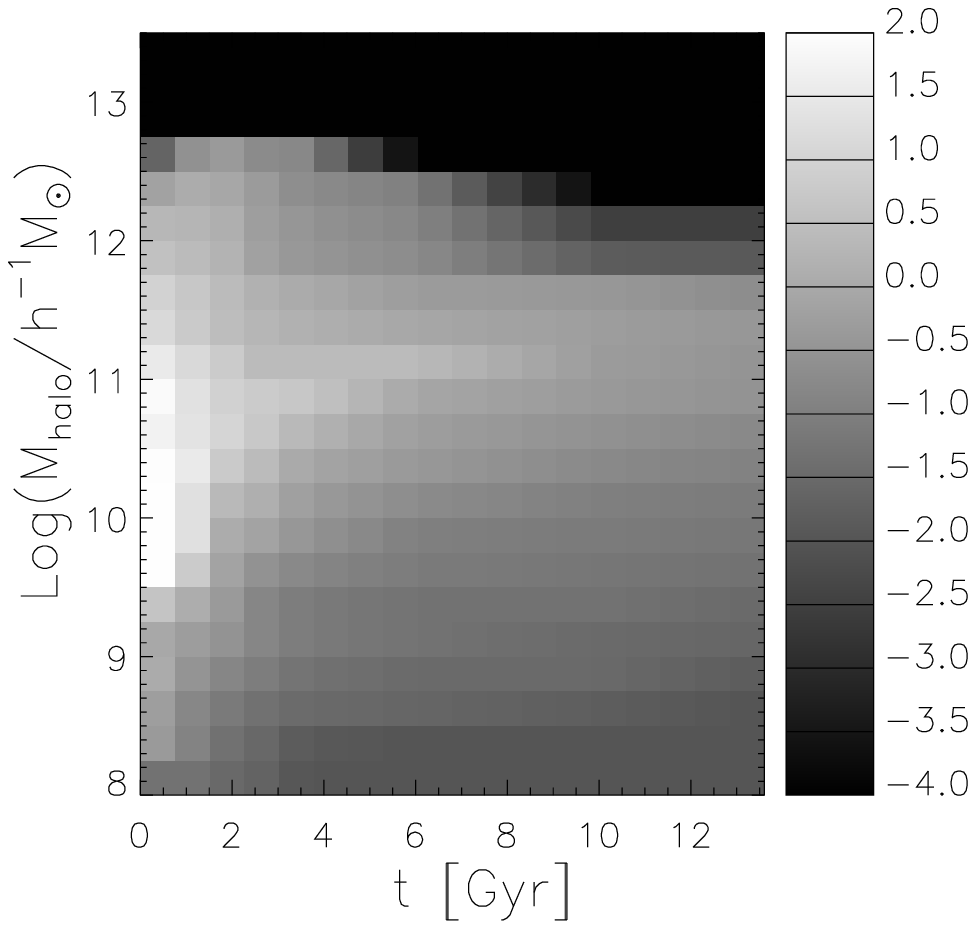}}\\%
\caption{Cooling efficiency as a function of halo mass and cosmic time in
model 4). The gray scale 
shows Log values of cooling efficiency in units of Log$[{\rm Gyr}^{-1}]$. 
X-axis is the cosmic time, with $13.7$ Gyr corresponding to present day. 
}
\label{fig:coolingmodel4}
\ec
\end{figure}

Fig.~\ref{fig:coolingmodel4} shows the cooling efficiencies used for 
model 4) as presented in Sec.2.6, and table \ref{tab:cooling4} lists the 
explicit values of those efficiencies.

\begin{table}
\caption{Values of cooling efficiencies in units of Gyr$^{-1}$. The values 
shown here are identical for models 0), 1), 2), 3) and 3b), and are plotted 
in Fig.~\ref{fig:cooling}. Halo mass is in units of $h^{-1}{\rm M}_{\odot}$ and 
is shown at the left column, time is in Gyr.}
\begin{center}
\begin{tabular}{l || c | c | c | c | c | c }
\hline
   Log$({\rm M}_{halo})$  & $t=0.80$  & 2.24 & 3.38 & 5.97 & 10.27 & 13.58     \\
       &  $z=7$  &  3.0 &   2.0  &  1.0 &   0.3  &   0  \\
\hline
  8.00 &  -1.30 &  -1.60 &  -2.00 &  -2.00 &  -2.00 &  -2.00 \\
  8.25 &  -0.90 &  -1.50 &  -1.90 &  -2.00 &  -2.00 &  -2.00 \\
  8.50 &  -0.80 &  -1.30 &  -1.50 &  -1.60 &  -1.80 &  -2.00 \\
  8.75 &  -0.50 &  -1.00 &  -1.30 &  -1.50 &  -1.50 &  -1.80 \\
  9.00 &  -0.30 &  -0.80 &  -1.10 &  -1.30 &  -1.50 &  -1.60 \\
  9.25 &   0.10 &  -0.80 &  -1.10 &  -1.30 &  -1.30 &  -1.50 \\
  9.50 &   0.80 &  -0.50 &  -0.80 &  -1.10 &  -1.20 &  -1.30 \\
  9.75 &   1.30 &  -0.10 &  -0.50 &  -1.00 &  -1.10 &  -1.20 \\
 10.00 &   1.30 &   0.20 &  -0.30 &  -0.70 &  -1.00 &  -1.10 \\
 10.25 &   1.53 &   0.50 &  -0.07 &  -0.40 &  -0.76 &  -0.93 \\
 10.50 &   1.36 &   0.81 &   0.29 &  -0.23 &  -0.57 &  -0.76 \\
 10.75 &   1.32 &   0.83 &   0.69 &  -0.06 &  -0.38 &  -0.53 \\
 11.00 &   1.13 &   0.42 &   0.42 &   0.41 &  -0.32 &  -0.49 \\
 11.25 &   0.77 &   0.33 &   0.17 &  -0.06 &  -0.28 &  -0.44 \\
 11.50 &   0.71 &   0.21 &   0.02 &  -0.32 &  -0.41 &  -0.67 \\
 11.75 &   0.43 &  -0.15 &  -0.45 &  -0.71 &  -0.83 &  -0.90 \\
 12.00 &   0.29 &  -0.23 &  -0.51 &  -0.73 &  -0.81 &  -0.92 \\
 12.25 &   0.07 &  -0.28 &  -0.51 &  -0.79 &  -1.50 &  -2.00 \\
 12.50 &  -0.58 &  -0.70 &  -0.80 &  -1.50 &  -2.00 &  -4.00 \\
 12.75 &  -0.78 &  -0.70 &  -1.50 &  -2.00 &  -4.00 &  -9.00 \\
 13.00 &  -1.58 &  -0.80 &  -1.80 &  -4.00 &  -4.00 &  -9.00 \\
 13.25 &  -4.00 &  -4.00 &  -4.00 &  -9.00 &  -9.00 &  -9.00 \\
\hline
\end{tabular}
\end{center}
\label{tab:cooling0}
\end{table}

\begin{table}
\caption{Same as table \ref{tab:cooling0}, but for model 4), and only showing
cooling efficiencies for haloes more massive than $10^{11.5}h^{-1}{\rm M}_{\odot}$.
For haloes less massive, the values are the same as in table \ref{tab:cooling0}.}
\begin{center}
\begin{tabular}{l || c | c | c | c | c | c }
\hline
   Log$({\rm M}_{halo})$  & $t=0.80$  & 2.24 & 3.38 & 5.97 & 10.27 & 13.58     \\
     &  $z=7$  &  3.0 &   2.0  &  1.0 &   0.3  &   0  \\
\hline
 11.50 &   0.71 &   0.21 &   0.02 &  -0.32 &  -0.41 &  -0.67 \\
 11.75 &   0.43 &  -0.15 &  -0.45 &  -0.71 &  -1.83 &  -1.90 \\
 12.00 &   0.29 &  -0.23 &  -0.51 &  -0.83 &  -2.51 &  -2.52 \\
 12.25 &   0.07 &  -0.28 &  -0.70 &  -1.00 &  -4.00 &  -4.00 \\
 12.50 &  -0.58 &  -0.70 &  -0.80 &  -4.00 &  -4.00 &  -4.00 \\
 12.75 &  -9.00 &  -9.00 &  -9.00 &  -9.00 &  -9.00 &  -9.00 \\
 13.00 &  -9.00 &  -9.00 &  -9.00 &  -9.00 &  -9.00 &  -9.00 \\
 13.25 &  -9.00 &  -9.00 &  -9.00 &  -9.00 &  -9.00 &  -9.00 \\
\hline
\end{tabular}
\end{center}
\label{tab:cooling4}
\end{table}
%

\end{document}